\documentclass{sigchi}

\CopyrightYear{2016}
\setcopyright{acmlicensed}
\doi{http://dx.doi.org/10.475/123_4}
\isbn{123-4567-24-567/08/06}
\conferenceinfo{CHI'16,}{May 07--12, 2016, San Jose, CA, USA}
\acmPrice{\$15.00}

\usepackage{balance}

\usepackage[bf,small]{caption}
\usepackage{subfig}
\usepackage{cite}
\usepackage{epsfig}
\usepackage{array}
\usepackage{color}
\usepackage{pifont}
\usepackage{multirow}
\usepackage{booktabs}
\usepackage{fixltx2e}
\usepackage[table]{xcolor}

\usepackage{epstopdf}

\usepackage{tikz}
\usepackage{amsmath}

\usepackage{url}

\usepackage{times}

\usepackage{color}

\def\plainkeywords{Search Bias; Search Bias Quantification; Sources of Search Bias; Social Media Search; Political Bias Inference; Twitter;}

\makeatletter
\def\url@leostyle{
  \@ifundefined{selectfont}{
    \def\UrlFont{\sf}
  }{
    \def\UrlFont{\small\bf\ttfamily}
  }}
\makeatother
\def\pprw{8.5in}
\def\pprh{11in}

\begin{document}

\CopyrightYear{2017}
\setcopyright{acmlicensed}
\conferenceinfo{CSCW '17,}{February 25-March 01, 2017, Portland, OR, USA}
\isbn{978-1-4503-4335-0/17/03}\acmPrice{\$15.00}
\doi{http://dx.doi.org/10.1145/2998181.2998321}

\title{Quantifying Search Bias: Investigating Sources of Bias\\ for Political Searches in Social Media}

\numberofauthors{7}
\author{
\alignauthor Juhi Kulshrestha\\
       \affaddr{MPI-SWS, Germany}
\alignauthor Motahhare Eslami\\
       \affaddr{University of Illinois at Urbana-Champaign, USA}
\alignauthor Johnnatan Messias\\
       \affaddr{MPI-SWS, Germany}
\alignauthor Muhammad Bilal Zafar\\
       \affaddr{MPI-SWS, Germany}\\
\and
\alignauthor Saptarshi Ghosh\\
       \affaddr{IIEST Shibpur, India}\\
\alignauthor Krishna P. Gummadi\\
       \affaddr{MPI-SWS, Germany}
\alignauthor Karrie Karahalios\\
       \affaddr{University of Illinois at Urbana-Champaign, USA}
              \affaddr{Adobe Research, USA}
}

\maketitle

\begin{abstract}

\noindent

Search systems in online social media sites are frequently used to find information about ongoing events and people. For topics with multiple competing perspectives, such as political events or political candidates, bias in the top ranked results significantly shapes public opinion. However, bias does not emerge from an algorithm alone. It is important to distinguish between the bias that arises from the data that serves as the input to the ranking system and the bias that arises from the ranking system itself. In this paper, we propose a framework to quantify these distinct biases and apply this framework to politics-related queries on Twitter. We found that both the input data and the ranking system contribute significantly to produce varying amounts of bias in the search results and in different ways. We discuss the consequences of these biases and possible mechanisms to signal this bias in social media search systems' interfaces.

\end{abstract}



\section{Introduction}  \label{sec:intro}

As algorithmic decision making systems pervade our daily lives, there
is a growing concern about their lack of transparency and their
potential for offering biased or discriminatory results to their
users~\cite{wh-study-on-biases}. Search engines are one of these
systems that users rely on for a variety of daily tasks from locating
specific websites or content to learning broadly about events, people,
or businesses unfamiliar to them.
When a user is trying to ``learn'' or gather information about a
topic~\cite{Welch-diversity-info-queries},
search engines could influence the user's opinions about the topic by
preferentially ranking results that correspond to one particular
perspective on the topic above others. This possibility is
particularly troubling in the context of informational queries about
polarizing topics like political debates or politicians, where
contrasting perspectives exist. Recently conducted field studies have
shown that not only do users place greater trust in higher ranked
search results ~\cite{google-we-trust}, but also that user opinions
about political candidates can be manipulated by biased rankings of
search results~\cite{search-engine-manipulation}.

Characterizing the bias present in a search system can be quite
challenging. Search engines employ a ranking system to retrieve a list
of results from a corpus of data (i.e., collection of web pages or
user posts like tweets) based on their relevance to a user's
query. So, when quantifying the bias of a search engine in response to
a query, it is useful to distinguish (i) the bias that arises from the
data corpus that acts as input to the ranking system of the search
engine, from (ii) the bias that arises from the ranking system itself.
In this paper, we shed light on these challenges by answering the
following research questions:

\begin{itemize}

\item RQ1: How can we quantify the different sources of search engine bias?
\item RQ2: How biased are the search results for political topics on social media sites like Twitter? Where does this bias in the search results come from?

\end{itemize}

We address these questions for political queries related to the 2016
US Presidential primaries using Twitter Search. We chose a social
media search engine because people are increasingly relying on social
media for their news. In recent elections, it has become common for
people to search for information about political candidates and
events on social media sites~\cite{Teevan_TwitterSearch}.
Furthermore, in August 2015, the Pew Research Center announced that
approximately two thirds of American adults received their news from
Twitter---and they continued to follow news threads on Twitter as they
progressed~\cite{twitter-pew}.

\noindent To answer our first question, we first propose
a search bias quantification framework that
quantifies the bias of the output results of a search engine. Moreover, this framework also discerns to what extent this output bias is due to the input data set that feeds into the ranking system and how much is due to the bias introduced by the ranking system.
Our framework requires a methodology for inferring political bias of individual search results. Hence, we also developed a method to infer the political bias of individual data items on Twitter with high accuracy and coverage.

To address the second question,
we gathered data for 25 political queries in Twitter search in December 2015 during a week in which two presidential debates occurred---one Republican debate and one Democratic debate.
Applying our quantification method on the collected data, we found that both the input data and the ranking system contribute significantly in producing the bias in the resulting Twitter search results. For example, while the Twitter stream containing our selected queries (the input data) contributed to the output data by adding a democratic bias on average,  the ranking system shifted or in some cases altered the polarity of this bias, resulting in a substantially different political bias in the final search results.

This collective contribution of the input data and the ranking system in producing the output bias can noticeably affect a user's search experience. We observed, for instance, that the Twitter input data stream for the most popular candidates in a party were more biased towards the opposing political perspective. However,  the  manner in which the ranking system altered the input to produce the output, and therefore its bias, was different for the popular candidates from each political party. While the ranking system mitigated the opposing bias in the search results for the most popular democratic candidate, it enhanced it for the most popular republican candidate. Simply put, if a user searches for the most popular republican candidate, she will get more tweets from the opposing political party than if she searched for the most popular democratic candidate. This may be less than desirable for a popular republican candidate if the users with the opposing polarity primarily post negative tweets about the candidate that result in negatively biased search results for her or him.
Additionally, we also observed that differently phrased but similar queries (referring to the same debate event) can yield significantly different biases.

Lastly, we call for raising the awareness of search engine users by signaling the bias in the search results. We briefly discuss two possible solutions of either incorporating the bias in the ranking systems itself or incorporating it in search engine designs by making the bias in the search results transparent to users.

\section{Related Work}  \label{sec:related}

\subsection{Bias in Web Search}
In recent years, there has been a growing interest in studying the bias in Web search engines results~\cite{sep-ethics-search,search-engine-bias-thesis,search-engine-bias-mitigation,Vaughan-search-coverage-bias,measure-search-engine-bias}. Much of this work has been motivated by the concern that dominant search engines like Google might favor certain websites over others when ranking relevant search results. For example, Google might
preferentially rank videos from YouTube (owned by Google) over videos from competing sites.
Whereas, exploring geographical bias, \cite{Vaughan-search-coverage-bias} measured whether sites from certain countries are covered more than sites from other countries in search results.

Several studies have explored political bias in Web search results and search
queries.  While Weber et al  \cite{Weber-query-logs} inferred political
leanings of search {\it queries} by linking the queries with political blogs,
Epstein and Robertson \mbox{\cite{search-engine-manipulation}} studied how bias in search rankings can influence people's candidate selection in elections. They asked people unaware of the political candidates in an election to search for the candidates and form an opinion based on the results. By biasing the search results in a controlled manner,
they showed that search rankings could impact their voting preferences in an election by $20\%$ or more. Their observations
 demonstrate the potential impact of search results on the political opinion of users, and thus provide the motivation for our present study.

A complementary line of work has studied the differences or biases in Web search results due to personalization, i.e., the differences in the
results seen by different users for the same query. \cite{web-search-personalization,web-search-personalization-geolocation}
studied differences in search results due to the geo-location of users.  \cite{Koutra-events-controversies} studied how the information-seeking behavior of users changes due to a major event such as a shooting incident, and found that most people use search engines to access information that they agree with.  These studies are performed from the viewpoint of particular users, whereas the present work shows that  political biases
exist even in {\it non-personalized} search results in the context of social media.

\subsection{Measuring Political Bias on Social Media}

While there have been prior attempts to infer the political bias of textual content in online forums, such as blogs and news articles~\cite{Adamic-blogosphere,biased-language-naacl,news-political-leaning-icwsm} or hashtags on Twitter \cite{weber2013political}, to our knowledge, no work has measured the political bias of a tweet based on its textual content, particularly due to its very small size.
Instead, many previous studies have tried to infer the bias of a tweet's source (the user who posted it),
by modeling the language usage of the users from different political polarities \cite{twitter-language-leaning-plosone,political-preference-twitter-asonam} or by leveraging the social links that a user has with other Twitter users with known biases~\cite{political-preference-twitter-sigchi,political-polarization-twitter-icwsm}. Conover {\it et al.}~\cite{political-alignment-twitter-socialcom} inferred the political alignment of Twitter users, based on  the tweets or hashtags posted by the users, as well as the community structure of diffusion networks (via retweets) of political information, observing that the mechanism relying on social network performs better. In \cite{impartiality-icwsm}, the authors automatically measure the impartiality of the social media posts based on how discernible the affiliation of the author is.

While all the aforementioned studies try to infer a Twitter user's bias, they rely on the assumption that the political leaning of a user is explicit in either her language or her social network. This may not be the case with many users who either do not post political content frequently or do not connect to other political users directly, but might still have political leanings. To account for this constraint, we build upon these prior studies to propose a new methodology to infer the bias of a user by including other factors correlated with the political affiliation of a user, namely her interests.

\subsection{Cross-Ideological Interactions on Social Media}

With the rising popularity of social media sites like Twitter and Facebook, users are increasingly relying on them to obtain news\mbox{\cite{neiman_news}}, real-time information about ongoing events and public opinion on celebrities\mbox{\cite{Teevan_TwitterSearch}}.
While some researchers
envisaged increased democratization through social media usage, with higher engagement
between users who do not share the same
political ideology \cite{semaan2014social},
some others argued that social media usage can result in
selective exposure by providing a platform that
reinforces users' existing biases \cite{liu2014twitter}.
By examining cross-ideological exposure through content and network analysis, \cite{himelboim2013birds} showed that political talk on Twitter is highly partisan and users are unlikely to be
exposed to cross-ideological content through their friendship network.
Other studies have also confirmed these results by demonstrating
users' higher willingness to communicate with other like-minded social media users
\cite{liu2014twitter, smith2013role} and their inability
 to engage in meaningful discussions with different-minded users \cite{yardi2010dynamic} .

To understand the political bias in social media better, many researchers have studied political polarization on Twitter through analyzing different groups' behavior. \cite{political-polarization-twitter-icwsm} showed that Twitter users usually retweet the users who have the same political ideology as themselves, making the retweeting network structure highly partitioned into left- and right-leaning groups with limited connections between them. Liao et al. \cite{liao2016snowden} investigated political polarization in the context of a majority and minority group on a political topic (pro-snowden vs. anti-snowden), finding that while the Twitter population is more likely to be pro-snowden, the minority group (anti-snowden) uses some features of the biased environment such as retweeting to mitigate this existing bias.

Instead of studying the interactions between cross-ideological users, our current study explores the problem of measuring political bias of social media search engine and unpacking the sources of bias.
While many studies have shown the existence of political bias in social media, to our knowledge none have characterized the sources of this bias. Little is known about how much bias is inherent in the data and whether and how the search system enhances or mitigates this bias. Our study tries to shed light on these questions through investigating the sources of political bias in the context of 2016 US presidential election on Twitter.

\subsection{Auditing Algorithms}
Today, algorithms that curate and present information in online platforms can affect users' experiences significantly -- creating discriminatory ads based on gender \cite{datta2014automated} or race \cite{sweeney2013discrimination}, showing different prices for the same products/services to different users \cite{hannak2014measuring} and mistakenly labeling a black man as an ape by an image tagging algorithm \cite{flicker-autotag} -- are some such examples. These issues have lead researchers, organizations and even governments towards a new avenue of research called ``auditing algorithms", which endeavors to understand if and how an algorithmic system can cause biases, particularly when they are misleading or discriminatory to users \cite{sandvig2014auditing,wh-study-on-biases}.

Besides the algorithm design, biased input data to an algorithm can also result in a biased output. This insight is particularly crucial in this digital era where many algorithms are trained using huge amounts of data \cite{barocas2014big}. Therefore, distinguishing whether a bias in an algorithmic system's output is caused by its input or the algorithm itself, is of prime importance. This work is a first step towards tackling this challenge in the area of search systems.

\section{RQ1: Quantifying Search Engine Bias} \label{sec:quantifying_bias}

\noindent This section focuses on our first research question -- the quantification of search bias.
In this paper, we quantify the search bias for the Twitter social media search engine
 in the context of the US political scenario, where there are two main
  political parties: the Democratic party and the Republican
party.

We first propose a framework to capture the different stages
of a search process and then discuss metrics to measure the biases at
each stage (RQ1a). This framework requires a methodology for inferring the
bias of individual search items, and later in the section we present such a methodology
in the context of Twitter search (RQ1b).

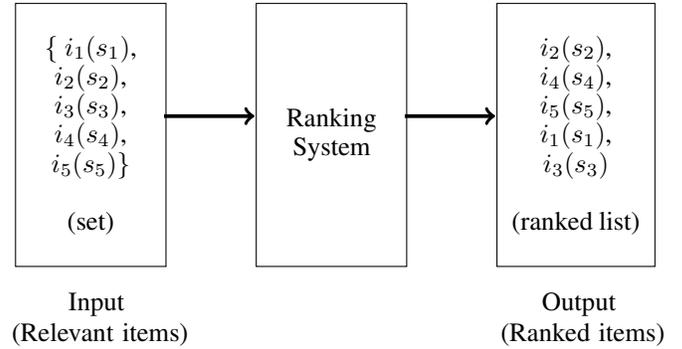
\begin{figure}[tb]
\center

\begin{tikzpicture}

\draw (0,0) rectangle (2,3.5) node[align=center,pos=.5] {
\{
$i_1 (s_1)$,\\
$i_2 (s_2)$,\\
$i_3 (s_3)$,\\
$i_4 (s_4)$,\\
$i_5 (s_5)$\}\\
\\
(set)
};
\node[text width=2.2cm] at (1.8,-0.5) {Input};
\node[text width=2.5cm] at (1.2,-0.9) {(Relevant items)};
\node (A) at (1.85,2) {};

\node (B) at (3.3,2) {};
\draw (3.2,0) rectangle (5.2,3.5) node[align=center,pos=.5] {
Ranking \\
System
};
\node[text width=2.2cm] at (4.6,-0.5) {};
\node (C) at (5.05,2) {};

\node (D) at (6.5,2) {};
\draw (6.4,0) rectangle (8.5,3.5) node[align=center,pos=.5] {
$i_2 (s_2)$,\\
$i_4 (s_4)$,\\
$i_5 (s_5)$,\\
$i_1 (s_1)$,\\
$i_3 (s_3)$\\
\\
(ranked list)
};
\node[text width=2cm] at (8.0,-0.5) {Output};
\node[text width=2.5cm] at (7.7,-0.9) {(Ranked items)};

\draw[-latex,line width=0.5mm][->] (A) -- (B) node [pos=0.5,above] {} node [pos=0.5,below] {} ;
\draw[-latex,line width=0.5mm][->] (C) -- (D) node [pos=0.5,above] {}node [pos=0.5,below] {} ; ;

\end{tikzpicture}

\caption{{\bf Overview of our search bias quantification framework.}
For a given query $q$, a set of data items relevant to the query is first selected.
Each individual data item (e.g., $i_1$, $i_2$) has some bias (e.g., $s_1$, $s_2$).
Then this set of relevant items is input to the ranking system which produces a ranked
list of the items. We define metrics for capturing the bias in the set of relevant items
input to the ranking system (input bias), and in the ranked list output by the ranking system (output bias).}
\label{fig:seach_bias_framework}
\end{figure}

\subsection{RQ1a: Search Engine Bias Quantification Framework}\label{sub:search-framework}

\noindent

Figure~\ref{fig:seach_bias_framework} gives an intuitive illustration of the main
stages of retrieving information via an algorithmic search system.
The search system retrieves information from a corpus of data items.
Each of these data items has
a bias associated with it, given by the bias
score $s$ (computed as described in the next section: RQ1b).
When a user submits a search query, a set of data items that is relevant to the query is first retrieved
(selected out of all data items in the entire corpus).
Then, the {\it set of retrieved relevant items} constitutes the input data to the
ranking system which produces a
{\it ranked list} of the relevant items, and
this ranked list is shown to the user as the search output.\footnote{The framework can also be generalized to modern-day IR systems which perform retrieval and ranking together, such as systems using topic modeling. We comment on this issue in the Discussion section.}
The users generally give much more attention and
importance to the top-ranked items in the list
than the lower-ranked items~\cite{irbook-manning}.

In the case of Twitter search engine, the data corpus comprises of the set of all
tweets. When a query like `Bernie Sanders' is issued to Twitter
search, all the tweets containing `Bernie Sanders' are selected
as relevant items. This set of relevant tweets serves as the input to the
ranking system, which orders them to output a ranked list of tweets
which are displayed to the user.

In our framework, we quantify three different biases for a
search system, in terms of the biases of the individual data items:
(i)~{\it input bias}: the bias in the set of retrieved items relevant to the query (filtered out of the whole corpus), that serves as the input data to the ranking system,
(ii)~{\it ranking bias}: the bias introduced by the
ranking system, and
(iii)~{\it output bias}: the cumulative bias in the ranked list output
by the search system.
In the remainder of the section, we discuss the metrics we use
to quantify these biases for Twitter search in the context of US politics.

\begin{table}
\renewcommand{\arraystretch}{1.3}
\center
\small
\begin{tabular}{|c|c|p{0.5\columnwidth}|}
\hline
Rank $r$ &  Bias till rank $r$ & Value \\ \hline
1 & $B(q, 1)$ & $s_2$  \\ \hline
2 & $B(q, 2)$ & $\frac{1}{2} (s_2 + s_4)$  \\ \hline
3 & $B(q, 3)$ & $\frac{1}{3} (s_2 + s_4 + s_5) $  \\ \hline
4 & $B(q, 4)$ & $\frac{1}{4} (s_2 + s_4 + s_5 + s_1)$  \\ \hline
5 & $B(q, 5)$ & $\frac{1}{5} (s_2 + s_4 + s_5 + s_1 + s_3)$  \\ \hline
\hline
\multicolumn{2}{|c|}{Output bias at rank 5} & $ \frac{1}{5} [ s_2( 1 + \frac{1}{2} +  \frac{1}{3} + \frac{1}{4} + \frac{1}{5} ) + \newline s_4 (\frac{1}{2} +  \frac{1}{3} + \frac{1}{4} + \frac{1}{5}) + \newline  s_5 ( \frac{1}{3} + \frac{1}{4} + \frac{1}{5} ) +  \newline  s_1 ( \frac{1}{4} + \frac{1}{5} ) + \newline  s_3 ( \frac{1}{5} ) ]  $ \\ \hline
\end{tabular}
\caption{{\bf Explaining the bias metrics with reference to Figure~\ref{fig:seach_bias_framework}.}}
\label{tab:bias-metrics}
\end{table}

\noindent {\bf Bias of an individual data item:}
As mentioned earlier, the search scenario that we are considering is one of the US politics, where there are
mainly two political parties.
Each data item (i.e., a tweet) can be positively biased (i.e., supporting) or negatively biased (i.e., opposing)
or neutral towards each of these two parties, and the bias score of each item (indicated by $s_i$ in Figure~\ref{fig:seach_bias_framework}) captures the degree to which the item is biased with respect to the two parties.
In the next section (RQ1b), we describe a methodology for measuring the bias score of items in the context of US political searches on Twitter social media.

We next define metrics for the input bias, output bias and ranking bias, in
terms of the bias scores of the individual data items.

\noindent {\bf Input Bias:}
When a user issues a query to the search system, a set of items that are relevant to the query is selected out of the whole corpus, and provided as input to the ranking system.
This input data captures the bias introduced by the query by filtering the relevant items from the whole corpus of data.
Therefore, we measure the input bias for a query as the aggregate bias of all items relevant to the query, that become the input to the ranking system.
Put differently, input bias gives a measure of what bias a user would have observed,
had she been shown {\it random} items relevant to the query, instead of a list ranked by the ranking system.

Specifically, the Input Bias $IB(q)$ for query $q$ is
the average bias of all $n$
data items that are
relevant to $q$
\begin{equation}
IB(q) = \frac{ \sum_{i=1}^{n} s_i  }{n}
\end{equation}
where the summation is over all the bias scores ($s_i$) of the $n$ data items found relevant to $q$.
For instance, for the query $q$ shown in Figure~\ref{fig:seach_bias_framework}, the input bias
is simply $IB(q) = \frac{1}{5} (s_1 + s_2 + s_3 + s_4 + s_5)$.

\noindent {\bf Output Bias:}
The output bias is the effective bias presented to the
user (who issued the search query) via the final ranked list from the search engine.
While quantifying search bias, it must be noted that, whereas the input bias was for an unordered set of items,
the output search bias is to be measured over a {\it ranked list}. The
higher ranked items should be given more importance, since not only are the users more
likely to browse through the top search results~\cite{irbook-manning}, but they also tend to have more trust in them~\cite{google-we-trust}.
Hence, we propose a metric for output search bias, that is inspired by the well-known metric Average Precision~\cite{irbook-manning}
from the Information Retrieval literature.

For a given search query $q$,
we first define the bias till a particular rank $r$ in the ranked results (i.e., the aggregate bias of the top $r$ results).
The bias $B(q, r)$ till rank $r$ of the
output ranked list is
defined as
\begin{equation}
B(q, r) = \frac{ \sum_{i=1}^{r} s_i }{r}
\label{}
\end{equation}
where the summation is over the top $r$ items in the ranked list.
For instance, with respect to the situation shown in Fig.~\ref{fig:seach_bias_framework},
Table~\ref{tab:bias-metrics} (first five rows) shows the bias till ranks 1, 2, ..., 5.

Next, we extend the above definition to define
the Output Search Bias $OB(q, r)$ for the query $q$ at rank $r$.
\begin{equation}
OB(q, r) =  \frac{ \sum_{i=1}^{r} B(q,i) }{ r}
\label{eq:OB}
\end{equation}
For instance, the last row of Table~\ref{tab:bias-metrics} computes $OB(q, r)$ at rank $r = 5$ with respect to the situation shown in Figure~\ref{fig:seach_bias_framework}.
Note that the bias score $s_2$ of the top-ranked item $i_2$ is given the highest weight, followed
by the bias score $s_4$ of the second-ranked item $i_4$, and so on.
This follows the intuition that bias in the higher ranked items are likely
to influence the user more than bias in the lower ranked items.\footnote{In case the bias score
of a particular item in the ranked list cannot be inferred,
this item can be ignored and the rankings can be re-computed. This is
similar to how missing relevance judgements are handled in the Information Retrieval
literature~\cite{average-prec-incomplete-judgement-cikm}.}

\noindent {\bf Ranking Bias:} The ranking bias is intended to capture
the {\it additional} bias introduced by the ranking system, over
the bias that was already present in the set of relevant items (i.e., relevant to $q$)
input to the ranking system.
If possible, ranking bias could be measured by auditing the exact ranking system being
deployed by the search engines. However, for any commercial search engine deployed
in the real-world, it is infeasible to know the internal details of the ranking system.
Hence, we view the ranking system as a ``black box'' where we only observe the
inputs and outputs.

Thus, we define the Ranking Bias $RB(q, r)$ for the query $q$ as simply the difference
between the output bias and the input bias for $q$ (as defined earlier).
\begin{equation}
RB(q, r) =  OB(q, r) - IB(q)
\end{equation}

\noindent {\bf Time-Averaged Bias:}
To capture the overall trend in the
bias, we collect multiple
snapshots of search results, compute the different bias metrics for
each snapshot, and then compute the time-averaged values of the aforementioned
metrics.
For instance we compute the time-averaged output search bias
$TOB(q,r)$ as the average of the $OB(q,r)$ (given by Equation
\ref{eq:OB}) values measured at various instants of time.  Similarly,
we define $TIB(q)$ and $TRB(q,r)$ as the time-averaged input bias and
time-averaged ranking bias for query $q$ respectively.

\subsection{RQ1b: Measuring the Political Bias of an Individual Twitter Search Result}

For applying our search bias quantification framework to Twitter search in our chosen
 context of US politics, we require
an automated method for inferring the political bias of an individual Twitter search result -- a tweet.
To measure the bias of a tweet, we have two options
(i)~we can measure the {\it source bias}, i.e. the bias of the author of the tweet, or (ii)~we can measure the {\it content bias}, i.e., the bias of the content of the tweet.
Especially, since the content of the tweet is quite short (only 140 characters),
attempting to judge the bias from the short content
may not always be accurate; therefore, in this paper we choose to use
the source bias to measure the political bias of an individual Twitter search result.

In the rest of this section, we start by present our methodology for inferring source bias of a tweet and then evaluate how well it works. Finally, we end with a brief analysis of how well the source bias and content bias match each other in practice in the context of political search queries.

{\bf Source Bias - Inferring Political Bias of Twitter Users}

It has been reported that people's political affiliation is correlated with their various
attributes.\footnote{\url{http://2012election.procon.org/view.resource.php?resourceID=004818}}
In this work, we leverage this
insight to infer political affiliations of users on Twitter based on their interests.
Our methodology for inferring the political bias of a given Twitter user $u$
is based on the following steps --
(i)~generating two representative sets of users who are {\it known} to have a democratic or republican bias,
(ii)~inferring the topical interests of $u$, and
(iii)~examining how closely $u$'s interests match with the interests
of the representative sets of democratic and republican users.

{\bf Generating representative sets of democratic and republican users:} The first step in our methodology involves getting
representative sets of democratic and republican users.
One option would be to
collect a dataset of users who report their political affiliations on Twitter.
However, such a dataset would suffer from self-reportage problem,
while also being biased towards the group of users who have self reported.
Instead, we use the methodology in~\cite{whoiswho_wosn,cognos_sigir}, which
infers the topical attributes of a user $v$
by mining the
Twitter Lists that other users have included $v$ in.
Thus, we rely on  what other people are reporting about a user, and not what
she is identifying herself as.
Using this method, we extracted a seed set 865 users who
have been labelled with
the topic ``democrats''  and a seed set of 1,348 users labelled with the topic ``republicans''.
These seed sets of users
include politicians (e.g., Steny Hoyer, Matt Blunt),
political organizations (e.g., DCCC, Homer Lkprt Tea-party)
as well as regular users.

{\bf Inferring topical interests of a user:} After obtaining the representative sets of democratic and republican users, the next step is to infer a user's interests on Twitter. To address this challenge,
we rely on the methodology in~\cite{user-interests-recsys}, which for a user $u$, returns a list of topics of interest of $u$
along with the frequency of each topic.
Here the frequency of a topic indicates the number of users
whom $u$ follows, who have been labelled with this topic using the aforementioned method in~\cite{whoiswho_wosn,cognos_sigir}.
For instance, if a user $u$ follows 3 users tagged with ``politics'' and 4 users
tagged with ``music'', then the returned list would be \{politics: 3, music: 4\}.
We convert this $<$ topic, frequency $>$ list into a
weighted $tf\_idf$ vector for user $u$, where the $idf$-s are computed
considering the interest lists of all the users in our dataset.
We refer to this $tf\_idf$ vector as
the {\it interest-vector} $I_u$ of the user $u$.
Specifically, if a user $u$ follows $f$ users on a particular topic,
then the corresponding entry in the interest-vector of $u$ is
computed using $tf = 1 + \mbox{log} f$ and $idf =  \mbox{log}\frac{N}{n}$,
where $N$ is the number of all users in our dataset,
and $n$ is the number of users who follow at least one user tagged with this topic.

Note that, if we are unable to get the followings of a user
(due to her account being protected, or her following no one),
or if she follows too few users (less than 10), we are not able to
infer her interests and hence political leaning.
However, the study~\cite{user-interests-recsys}
showed that this methodology of inferring interests is applicable for
a very large fraction of active users in Twitter.

{\bf Matching interests of user to interests of democrats and republicans:} Using the aforementioned formulation, we also obtain the interest vectors
of all the users in the two seed sets.
Then, we construct a normalized aggregate interest vector for the democrat seed set ($I_D$) and
the republican seed set ($I_R$),
by aggregating the interest vectors of all users in the set and
normalizing such that each aggregate vector sums up to 1.0.
Some of the top terms in
$I_D$ are {\it [progressive, democrats, obama, dems, policy, liberals, p2, activists, dc, international]}
 while those in $I_R$ are
{\it [patriots, conservative, tcot, right, gop, republican, tea party, republicans, fox news, pundits]},
where the terms are ranked in decreasing order of their $tf\_idf$ scores.
We observe that apart from terms related to politics, the two vectors
also differ in terms of their emphasis on other topics, for example $I_R$
gives higher weight to sports-related terms, while $I_D$ gives
higher weight to terms related to technology and entertainment.
Thus, these vectors can be used to infer the likely political bias of
a wide range of users, even if they are not explicitly following politicians on Twitter, or even
if they are following politicians from both parties.

Finally, given a user $u$ whose political bias is to be inferred,
we obtain the interest-vector $I_u$ of $u$ (as described above), and
compute the cosine similarity of $I_u$ with $I_D$ and $I_R$.
Then the {\it bias score} of user $u$ is given by,
\begin{equation}
Bias(u) =  cos\_sim(I_u, I_D) - cos\_sim(I_u, I_R).
\end{equation}
We max-min normalize these differences in similarity, such that the bias score
of a user lies in the range $[-1.0, 1.0]$.
A bias score closer to $+1.0$ indicates that $u$ is more democratic leaning,
 and a score closer to $-1.0$ indicates that $u$ is more republican leaning.

\noindent {\bf Public deployment of the source bias inference methodology:}
We have publicly deployed the aforementioned source bias inference methodology in the form of a Twitter application, at
\url{http://twitter-app.mpi-sws.org/search-political-bias-of-users/}.
One can login to the application using her Twitter credentials, and see their inferred political affiliation.
One can also search for other Twitter users to check out their inferred political leaning.

{\bf Evaluation of Political Bias Inference Methodology}

To check whether the source bias inference methodology works well for whole
spectrum of politically interested
users,
we carry out the evaluation considering three test sets of Twitter users --
(i)~politically interested common users, selected randomly from the set of users who
have retweeted the two parties' accounts on Twitter,
(ii)~the current US senators, for whom it is well known which political
party they represent, and
(iii)~self-identified common users, each having fewer than 1000 followers, who have
themselves indicated their political ideology in their Twitter account bios.

We first use the set of politically interested common users to evaluate how good
our inferred bias score is.
After establishing that our bias score works, we define thresholds on the score
to categorize users into three distinct categories -- republican, neutral, democratic.
Finally we present how well our inference
methodology works for the senators and self identified common users.

We evaluate the methodology with respect to two metrics:
(i)~{\it coverage} -- for what fraction of Twitter users can the methodology infer the political bias, and
(ii)~{\it accuracy} -- for what fraction of users is the inference correct.

{\bf \textit{Evaluation for politically interested common users}}

To identify a set of politically interested common users, we followed the methodology
given in~\cite{liu2014twitter}.
We used the Twitter accounts of democratic (@TheDemocrats)
and republican (@GOP) parties as seed accounts,
and collected up to 100 retweeters of each of the most recent 3,200 tweets posted
by the two accounts.
Doing so, we collected 98,955 distinct users who retweeted @TheDemocrats and
71,270 distinct users who retweeted @GOP.
Out of these users, we randomly selected 100 retweeters
each of the democratic and republican account, giving us a test set of
200 common users who are politically interested.

{\bf Judging the ground truth bias of test users:}
To get the bias annotations for political leanings of these 200 users,
we conducted an Amazon Mechanical Turk (AMT) survey where human workers
were shown a link to a user's Twitter profile, and asked to
label the user as either pro-democratic, pro-republican, or neutral,
based on the user's profile attributes and the tweets posted by the user.
We got judgements from 50 distinct AMT workers for each test user.
The AMT bias score of each user was computed
by adding $+1$ for each pro-democratic judgement, $-1$ for each pro-republican judgement
and $0$ for each neutral judgement,
and then normalizing by the total number of judgements.
Thus, the AMT bias score for a user lies in the range $[-1.0, 1.0]$, where a more positive score
indicates a stronger democratic bias (with more AMT workers labelling the user as pro-democratic), whereas
a more negative score indicates a stronger republican bias (with more AMT workers
labelling the user as pro-republican).

{\bf Evaluating our inferred score:}
Our methodology could infer the bias of all the 200 selected users (coverage = 100\%).
To quantify the accuracy of the methodology,
we check whether our inferred bias scores correlate well with the AMT bias scores.

To verify this, we present Table~\ref{tab:AMT_bias_score_vs_Inf_bias_score1} and Table~\ref{tab:AMT_bias_score_vs_Inf_bias_score2}.
In Table~\ref{tab:AMT_bias_score_vs_Inf_bias_score1}, we bin our inferred bias scores for these test users
into three ranges [-1.0, -0.5], (-0.5, 0.5) and [0.5, 1.0], and then we compute the average AMT bias scores for users in each bin.
As can be seen from Table~\ref{tab:AMT_bias_score_vs_Inf_bias_score1},
the average AMT bias score for Bin 1 (corresponding to users inferred to be strongly republican) is strongly
republican leaning ($-0.86$), while the average AMT bias score for Bin 3 (corresponding to users inferred to be strongly democratic) is strongly democratic leaning ($0.93$).
Similar results are seen when we bin the users based on AMT bias scores and then
compute the average inferred bias score for the users in each bin (as shown in Table~\ref{tab:AMT_bias_score_vs_Inf_bias_score2}),
demonstrating that our inferred bias scores are well correlated with the AMT bias scores.

\begin{table}
\center
\small
\begin{tabular}{ | l | c |}
\hline
 {\bf Inferred bias score bins } & {\bf Avg. AMT bias score} \\
\hline
Inferred Bin 1 [-1.0, -0.5]  & $-0.86$ \\
\hline
Inferred Bin 2 (-0.5, 0.5) & 0.14\\
\hline
Inferred Bin 3 [0.5, 1.0] & 0.93\\
\hline
\end{tabular}
\caption{{\bf Match between the AMT bias score and our Inferred bias score: Average AMT bias scores of users binned according to their Inferred bias score.}}
\label{tab:AMT_bias_score_vs_Inf_bias_score1}
\end{table}

\begin{table}
\center
\small
\begin{tabular}{ | l | c |}
\hline
{\bf AMT bias score bins} & {\bf Avg. Inferred bias score} \\
\hline
AMT Bin 1 [-1.0, -0.5] & $-0.32$ \\
\hline
AMT Bin 2 (-0.5, 0.5) & $-0.02$ \\
\hline
AMT Bin 3 [0.5, 1.0] & 0.14\\
\hline
\end{tabular}
\caption{{\bf Match between the AMT bias score and our Inferred bias score: Average Inferred bias scores of users binned according to their AMT bias score.}}
\label{tab:AMT_bias_score_vs_Inf_bias_score2}
\end{table}

\begin{figure}[tb]
\center
\includegraphics[height=0.95\columnwidth,angle=-90]{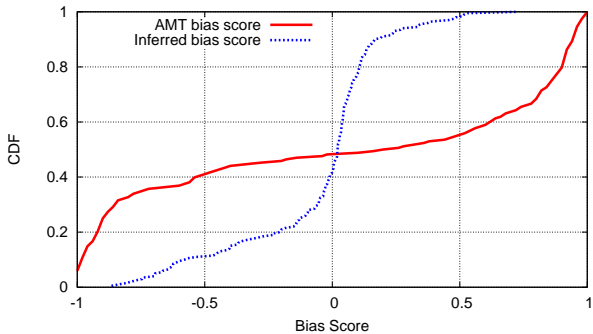}
\caption{{\bf CDF of AMT bias scores and Inferred bias scores for politically interested common users.}}
\label{fig:amt_bias_scores_inf_bias_scores_cdf}
\end{figure}

We also observe that, though there is high correlation between the
AMT bias scores and our inferred scores,
the spread of the distribution of the two scores in the interval $[-1.0, 1.0]$
are quite different. This difference is shown
in Figure~\ref{fig:amt_bias_scores_inf_bias_scores_cdf} which plots
the CDF of the two scores. We observe that while many of the AMT bias scores are
close to the boundaries of the interval, most of the inferred scores lie within $[-0.5, +0.5]$.
This difference in the spread of the distributions motivated us to
{\it discretize} our inferred bias score, and categorize users
into three classes -- democratic-leaning, republican-leaning, or neutral.

{\bf Discretizing the bias score into categories:}
Having established that our inferred bias score correlates well with the AMT bias score,
we now want to categorize users according to their bias.
To do so, we take the conservative approach of labelling a user as democratic or republican
only when we have high confidence, otherwise to label them as neutral.
To achieve this categorisation, we decided to fix a suitable threshold $x \in [0.0, 1.0]$, and label the users with inferred bias
score lying within $(-x, x)$ as neutral, the users having inferred bias score in the range $[-1.0, -x]$ as republicans,
and the users having inferred bias score in the range $[x, 1.0]$  as democrats.
We experimented with several choices for the threshold $x$ = 0.01, 0.03, 0.05, 0.08 and 0.1.
For each choice, we computed a {\it confusion matrix} of the match between AMT bias scores and the inferred bias scores.
For instance, Table~\ref{tab:AMT_bias_score_vs_Inf_bias_score-confusion-matrix} depicts the confusion matrix of the match
between AMT bias scores and the inferred bias scores for $x = 0.03$.
We found that the threshold of $x = 0.03$ maximizes the sum of the diagonal of the confusion matrix.
In the later sections of the paper,
we will label users as democrats or republicans, only when their inferred score falls outside
of the neutral zone $(-0.03, 0.03)$.

Inclusion of a neutral zone has both pros and cons. While now we have higher confidence in our labels of democrats and republicans,
 some of the users who are marginally democratic or republican may now get labelled as neutral.
 But we make this conservative choice for our analyses, so as to not overestimate the bias in the search results.

\begin{table}
\center
\small
\begin{tabular}{ | l | c | c | c | }
\hline
AMT bias score &  {\bf Inferred Rep} & {\bf Inferred Neutral} & {\bf Inferred Dem} \\
\hline
AMT Bin 1 & 84.05\% & 13.04\% & 2.89\% \\
\hline
AMT Bin 2 & 18.18\% & 45.45\% & 36.36\% \\
\hline
AMT Bin 3 & 3.89\% & 12.98\% & 83.11\% \\
\hline
\end{tabular}
\caption{{\bf Confusion matrix of the match between AMT bias scores and Inferred bias scores}}
\label{tab:AMT_bias_score_vs_Inf_bias_score-confusion-matrix}
\end{table}

\begin{table}[tb]
\center
\small
\begin{tabular}{ | l | l | l |}
\hline
{\bf Political Bias}  & {\bf Coverage}  & {\bf Accuracy} \\
\hline
\multicolumn{3}{|c|}{{\bf Current US Senators}} \\
\hline
Democratic (n=45) &  97.78\%  &  86.36\%   \\
Republican (n=54) &  98.15\%  &  98.11\% \\
\hline
Average &  97.96\%  &  92.23\% \\
\hline
\hline
\multicolumn{3}{|c|}{{\bf Self-identified common users}} \\
\hline
Democratic (n=426) &  92.01\%   &  88.52\%  \\
Republican (n=675) & 90.22\%    & 82.95\%    \\
\hline
Average &  91.12\%  &  85.73\% \\
\hline
\end{tabular}
\caption{{\bf Coverage and accuracy of the political bias inference methodology for (i)~current US senators, and (ii)~common users who have declared their political ideology in their Twitter account profile.}}
\label{tab:senators_common_coverage_accuracy}
\end{table}

{\bf \textit{Evaluation for popular users (US senators)}}

\noindent
The performance of the methodology for the 100 current US senators
is summarized in Table~\ref{tab:senators_common_coverage_accuracy}.
The methodology has a very high coverage for the US senators,
and failed to infer the political bias for only two senators, one from each party.
Further investigation showed that one of these senators
did not follow anyone on Twitter, while the other followed only one other user;
hence, we could not infer their topical interests.
In terms of accuracy, our methodology correctly identified the political bias of 98.1\% of the Republican senators
for whom bias could be inferred, and 86.4\% of the Democrat senators whose bias could be inferred.

{\bf \textit{Evaluation for self-identified common users:}}

For our final test set, we chose users who have themselves indicated their political ideology in their Twitter account biography.
Using the service Followerwonk (\url{http://moz.com/followerwonk}), we obtained users located in
 the United States, with fewer than 1000 followers
(to ensure that we get common users),
whose bios contained certain keywords matching democrats
(``democrat'', ``liberal'', ``progressive'') and
republicans (``republican'', ``conservative'', ``libertarian'', ``tea party'').
The bio of each user was then manually inspected, and we retained only those users
whose bios actually reflected their political ideology. For instance, users having bios like
``{\it I am a \#conservative \#Christian who is neither a \#Democrat nor a \#Republican, but an \#Independent voter}''
and ``{\it We hate Politicians - Democrats, Republicans, all of them.}'' were discarded.
Finally we obtained 426 self-identified democratic users and 675 self-identified republican users.

The performance of the proposed bias detection methodology on these users with self-identified biases
is also indicated in Table~\ref{tab:senators_common_coverage_accuracy}.
The coverage is 91.12\% on average, across democratic and republican users. Closer inspection of the
users for whom we could not infer the bias revealed that they were either protected accounts,
 or they followed too few accounts, and as a result their interests could not be inferred.

 The interest vectors of correctly inferred self-identified democratic users contain political terms like ``liberal'', ``progressive'',
 and ``dem'', as well as other terms like ``gay'', ``lgbt'', ``science'',
and ``tech''. In contrast, those for self-identified republican users contain
political terms like ``tea'', ``gop'', and ``palin'' along with other
related terms like ``patriots'', ``military'', and ``vets''.
On the other hand,
many of the users for whom we inferred the incorrect leaning either
have opposite leaning tags in their interest vectors, or their non-political
 interests end up matching the interests of the opposite side more than their own.
However, the overall accuracy of bias inference for these self-identified common users is also high
(85.73\% on average across democratic and republican users),
as shown in Table~\ref{tab:senators_common_coverage_accuracy}.

{\bf Match between Source Bias \& Content Bias}  \label{sec:source-content-bias}

\noindent
As stated earlier, we use the source bias (i.e., the bias of the user who posted a tweet)
to quantify the bias of a tweet, instead of using content bias.
In this section, we investigate {\it how closely source bias and content bias of a tweet
 reflect each other}.

\noindent{\bf Judging the content bias:}
For each of our selected queries like ``democratic debate'', ``republican debate'', ``hillary clinton'' and ``donald trump'' (more details on query selection in the next section),
we considered two Twitter search snapshots -- one during the republican debate
on December 15, 2015 and another during the
democratic debate on December 19, 2015.
In each snapshot we collected the first page (20 tweets) of search results, leading to a
total of 881 distinct tweets, which we use to analyze the extent to which source bias
and content bias match each other.

To get the content bias annotations for these 881 tweets,
we conducted an Amazon Mechanical Turk (AMT) survey where human workers
were shown each tweet (but {\it not} the user who posted it),
and they were asked
to label the tweet as pro-democratic, pro-republican, or neutral.
We got judgements from
10 distinct workers for each tweet.
The content bias score of each tweet was computed in the same manner as described in the earlier section titled ``Evaluation for politically interested common users''.

\definecolor{LightCyan}{rgb}{0.88,1,1}
\definecolor{LightRed}{rgb}{0.95,0.6,0.6}
\definecolor{LightGreen}{rgb}{1,1,0.88}
\definecolor{LightYellow}{rgb}{0.6,0.7,0.6}

\begin{table}
\center
\small
\begin{tabular}{ | c | c || c | c | c |}
\hline
{\bf Content Bias}  & {\bf Fraction}  & \multicolumn{3}{c|}{\bf Source Bias} \\ \cline{3-5}
  & {\bf of tweets}  & {\bf Frac} & {\bf Frac} & {\bf Frac} \\
    & {\bf }  & {\bf Dem} & {\bf Rep} & {\bf Neu} \\
\hline
\rowcolor{LightRed}
{\bf Strongly rep}  & 13.85\% & 74.59\% & 10.66\% & 14.75\%   \\
\rowcolor{LightRed}
$[-1.0, -0.75)$  & & & &   \\
\hline
\rowcolor{LightGreen}
{\bf Moderately rep}& 22.02\%  & 55.73\% & 13.02\% & 31.25\% \\
\rowcolor{LightGreen}
$[-0.75, -0.25)$   & & & &   \\
\hline
\rowcolor{LightCyan}
{\bf Weakly rep}& 10.33\% & 22.22\% & 22.22\%  & 55.56\%\\
\rowcolor{LightCyan}
$[-0.25, 0.0)$   & & & &   \\
\hline
\rowcolor{LightYellow}
{\bf Neutral} & 5.68\%  & 20.41\% & 8.16\% & 71.43\% \\
\rowcolor{LightYellow}
$[0.0,0.0]$   & & & &   \\
\hline
\rowcolor{LightCyan}
{\bf Weakly dem} &  10.44\%   &  23.33\% & 13.33\% & 63.33\% \\
\rowcolor{LightCyan}
$(0.0, 0.25]$   & & & &   \\
\hline
\rowcolor{LightGreen}
{\bf Moderately dem} &  23.16\%   &  16.5\%  & 16.0\% & 67.5\% \\
\rowcolor{LightGreen}
$(0.25, 0.75]$   & & & &   \\
\hline
\rowcolor{LightRed}
{\bf Strongly dem} & 14.53\%   &  9.76\%  & 9.76\% & 80.49\%\\
\rowcolor{LightRed}
$(0.75, 1.0]$  & & & &   \\
\hline
\end{tabular}
\caption{{\bf Fraction of tweets in the different ranges of the content bias score (based on AMT workers' judgement), and the match between the source and content bias in the different ranges.}}
\label{tab:bias_score_fraction_accuracy}
\end{table}

\noindent{\bf To what extent do source bias and content bias match each other?}
For studying the match between source bias and content bias, we divide the range of content bias scores
into 7 bins, as shown in Table~\ref{tab:bias_score_fraction_accuracy},
varying from neutral to strongly biased on both sides.
The first thing to note from Table~\ref{tab:bias_score_fraction_accuracy}
is that
when the content is strongly biased, the match between the source and content bias is high (about 75\% or more),
irrespective of whether the content is biased towards the democrat or the republican perspective.
This indicates that strongly biased content
is mostly produced by the users with the same source bias.
On the other hand, when the content is weakly biased, our inferred source bias
has little or no correlation with the
content bias, since such weakly biased / neutral content
is equally likely to be produced by users with either bias.

Next we present the results of applying our bias quantification framework, using the source bias inference methodology, for analyzing political searches on Twitter.

\section{RQ2: Characterizing Political Bias in Social Media Search}

To characterize the bias in Twitter search results via our framework, we first describe the queries we selected and the data we gathered from Twitter for our chosen context of US presidential primaries. We then analyze the collected data to understand the possible sources of bias in Twitter Search (RQ2a) and the interplay between the input data and the ranking system that produces the output bias -- the bias in the search results observed by Twitter users (RQ2b).

\noindent
\textbf{The Selection of Search Queries:} In order to study search bias via the proposed framework, it would
be ideal to have access to the actual search queries
that people are using on Twitter to get information about the
2016 US Presidential primary debates.
However, as researchers we did not have access to this proprietary dataset.
We, therefore,  followed the two step methodology used in~\cite{Koutra-events-controversies} of first identifying a seed set of queries and then expanding them to satisfy two properties.
Our seed set of queries is comprised of  the terms {\it ``democratic debate''},
and {\it ``republican debate''}, and their shortened versions
\textit{``dem debate''} and \textit{``rep debate''} that are popular
on Twitter to accommodate the short length of tweets.

Next we expanded our query set to include other likely related queries that
satisfied two properties: {(i) the queries should be used by many users, and
(ii) the queries should not be biased to a particular party.
To satisfy the first property, we used hashtags to expand our query set, since hashtags are used extensively on Twitter by users to tag and follow
discussions about political topics~\cite{political-alignment-twitter-socialcom}.
Moreover, hashtags act as recommended queries on Twitter; when a user clicks on a hashtag, a Twitter search page for that hashtag is presented to her.
To identify such hashtags, we collected the Twitter search results
for our four seed queries during the November 2015 republican
and democratic debates. We then identified the most frequently occurring
hashtags
containing the term ``debate'' to ensure that they are related to the presidential debates.
We found 57 hashtags related to the democratic debate and 63 related to the republican debate.
From these, we selected the top 10 hashtags popular within each political party. This resulted in 15 distinct hashtags.

In addition to being popular, we wanted the queries in our expanded set to be unbiased to avoid over estimating the bias in search results.
We, therefore,  removed biased search queries (e.g.,  \#debatewithbernie, \#hillarycantdebate), and only retained queries for whom its difficult to estimate the political leaning of the poster.  This approach resulted in the following queries: democratic debate, dem debate, \#democraticdebate, \#demdebate, republican debate, rep debate, \#republicandebate and \#gopdebate.
In addition, given that politically interested users (irrespective of their leaning) can search for any candidate by name, we also included the names of the 17 presidential candidates to our set of queries.
We use our quantification framework with these 25 queries to measure the bias for political searches on Twitter.

\textbf{Data Collection:} For our analysis, we selected a one week time period that contained both a republican and a democratic debate -- December 14 - 21, 2015.  The Republican debate aired on December 15, and the Democratic debate aired on December 19. Our goal was to capture a representative collection of tweets from Twitter Search that represented both parties at similar points in time and for a similar event.  From this time period, we collected Twitter search results for each of the selected queries.

Twitter search provides different filters on the search results including ``top'', ``live'', ``news'', ``photos'' and ``videos''.\footnote{\url{https://twitter.com/search-home}} The default,  ``top'' results, contain the tweets chosen by the proprietary Twitter ranking system based on many factors, including the number of users that engaged with a tweet\footnote{\url{https://support.twitter.com/articles/131209}}. This filter provides us with the output set for a given search query.  To collect the output set, we
collected the top 20 search results on the first page for each query at 10-minute intervals for the whole aforementioned time period. The political bias of this set is the \textit{Output Bias} defined in our quantification framework.
 In total, across all the selected queries, we collected 28,800 snapshots that included 34,904 distinct tweets made by 17,624 distinct users.

To collect the input data to the ranking algorithm, we used Twitter's streaming API to collect all of the tweets that contained our selected queries during our data collection period and used this chronologically ordered tweet stream to calculate the {\it Input Bias}. Overall, across all the queries, we collected more than 8.2 million tweets posted during this time period, by 1.88 million distinct users which comprise of the input data for our selected queries.

\textit{De-Personalizing the Search Results}: Our objective is to investigate possible bias created by the ranking of presented tweets.  Because bias may be introduced by personalization features associated with the Twitter user issuing the query, we focused on the consistent, non-personalized ranking of results shown to every user to examine bias. To mitigate personalization effects (e.g., geographical personalization based on IP addresses), all search queries were made without logging in to Twitter, and from the same IP subnet.

\begin{table}
\center
\small
\begin{tabular}{| l | c | c | c |}
\hline
{\bf Query}  & {\bf Output Bias}   &  {\bf Input Bias}   &  {\bf Ranking Bias} \\
				 &  {\bf (TOB)}   &  {\bf (TIB)}   &  {\bf (TRB)} \\
\hline
\multicolumn{4}{|c|}{{\bf Queries Related to Democratic Candidates}} \\
\hline
Hillary Clinton & 0.21 & 0.03  & 0.18 \\
Bernie Sanders &  0.71 & 0.55 & 0.16  \\
Martin O'Malley &  0.64 & 0.57  & 0.07  \\
\hline
\textbf{Average} &  $0.52$  &  $0.38$  &  $0.14$ \\
\hline
\multicolumn{4}{|c|}{{\bf Queries Related to Republican Candidates}} \\
\hline
Donald Trump &  0.29  &  0.19 &  0.10  \\
Ted Cruz &  $-0.48$  &  $-0.11$  & $-0.37$  \\
Marco Rubio &  $-0.41$  &  $-0.12$  & $-0.29$  \\
Ben Carson &  0.46 &  0.20  &  0.26 \\
Chris Christie &  $-0.14$ &  0.27   &  $-0.41$ \\
Jeb Bush &  $-0.31$  &   0.09  &  $-0.40$ \\
Rand Paul &  $-0.37$  &  $-0.18$  &  $-0.19$ \\
Carly Fiorina &  0.16  &   0.38  &  $-0.22$ \\
John Kasich &  $-0.09$  &  $-0.13$  &  0.04 \\
Mike Huckabee &  0.30  &  0.12  &  0.18 \\
Rick Santorum &  $-0.04$  &  0.18  &  $-0.22$ \\
Lindsey Graham &  $-0.45$  &  0.07  &  $-0.52$ \\
George Pataki &  $-0.17$  &  0.09  &  $-0.26$ \\
Jim Gilmore &  $-0.35$  &  $-0.11$  &  $-0.24$ \\
\hline
\textbf{Average} &  $-0.11$  &  $0.07$  &  $-0.18$ \\
\hline
\multicolumn{4}{|c|}{{\bf Queries related to democratic debate}} \\
\hline
democratic debate &  0.43 &  0.38  &  0.05 \\
dem debate &  0.52  &  0.29  & 0.23  \\
\#democraticdebate &  0.28  &   0.19  & 0.07 \\
\#demdebate &  0.57  &  0.56  & 0.01  \\
\hline
{\bf Average} &  0.45  &  0.35  & $0.10$  \\
\hline
\multicolumn{4}{|c|}{{\bf Queries related to republican debate}} \\
\hline
republican debate &  0.53  & 0.27 &  0.26 \\
rep debate &  0.31  &  0.40  &  $-0.09$ \\
\#republicandebate &  0.39  &  0.34 & 0.05  \\
\#gopdebate &  0.04 &   0.10  & $-0.06$  \\
\hline
{\bf Average} &  0.32  &  0.28  & $0.04$  \\
\hline
\end{tabular}
\caption{{\bf Time averaged bias in Twitter search ``top'' results, for selected queries (related to political candidates and debates) -- output bias $TOB$, input bias $TIB$, and ranking bias $TRB$. Here a bias value closer to $+1.0$ indicates democratic bias and a value closer to $-1.0$ indicates republican bias.}}
\label{table:query-topic-bias}
\end{table}

\subsection{RQ2a: Where Does the Bias Come from?}
\vspace{0.2cm}

{\bf \textit{It is Not Always the Ranking System: Input Data Matters}}

Table~\ref{table:query-topic-bias} shows the three biases (output, input, and ranking) for the selected queries. It reveals that queries related to democratic and republican candidates as well as democratic and republican debates have a democratic-leaning input bias (larger than 0) on average.
This observation implies that the full tweet stream containing these query-terms, without any interference from the ranking system, contains a more democratic slant -- although, the democratic bias for queries related to the republican debate and candidates is lower than that related to the democratic debate and candidates, on average.

It is important to note that there exists a pre-existing bias in the  input data to the ranking system, and that this input bias can have a significant effect on the final output search bias seen by the end user.
In the case of Bernie Sanders, for example, the output search bias is very democratic ($0.71$); only a small fraction of this bias comes from the Twitter ranking system, while the majority originates from the input data, indicating that most of the Twitter population that discusses Bernie Sanders on Twitter has a democratic leaning.
The effect of input data on the search results' bias highlights the importance of incorporating the input data when auditing algorithms, and teasing out how much of the bias is present in the data itself and how much is contributed by the algorithmic system.

The democratic leaning input bias on Twitter for a large majority of the queries
can be explained by the {\it bias of overall Twitter corpus}.
We measure the bias of the Twitter corpus in two ways:
(i)~{\it User population bias}: The bias of 1000 Twitter users selected randomly from the Twitter userid space (i.e., the user-ids were randomly selected
from the range of 1 through the id assigned to a newly
created account in December 2015), and
(ii)~{\it Full tweet stream bias}: The bias of 1000 tweets selected randomly from Twitter's 1\% random sample for December 2015.
For measuring the corpus bias in both cases, we applied the same methodology as used for measuring the input bias, i.e., measuring the bias of the users
and the source bias of the tweets.
These two approaches resulted in a population bias of $0.25$ and a full tweet stream bias of $0.3$.
These positive bias values show that the population of Twitter users is democratic-leaning, and this bias is even more democratic-leaning when we consider the active users whose tweets have been included in the full tweet stream (measured using Twitter's 1\% random sample). These findings are in-line with prior studies \cite{pew-twitter-dem} which have shown that Twitter has a high fraction of democratic leaning users.

It is important to note that, though the Twitter corpus has a democratic-leaning bias, the input bias (TIB) of the different queries varies across the spectrum (as shown in Table~\ref{table:query-topic-bias}).
This variation is because each query acts as a filter to extract a subset of Twitter users whose tweets are relevant to that queries, and the sets of users filtered out by different queries have differing biases.
Therefore, athough the corpus bias of Twitter is uniform, the specific query
being considered determines the input data set and hence the input bias, which in turn affects the final
output bias observed by the user.

{\bf \textit{The Power of the Ranking System}}

The ranking biases in Table~\ref{table:query-topic-bias} show that while data has a major role in the search results' bias, the ranking system still plays a significant role by shifting the bias or even changing its polarity. Comparing the output bias and the input bias for the democratic and republican candidate queries reveals an interesting discovery---on average, the ranking system shifts the bias of each category towards that party's leaning.
For democratic candidate queries, on average, the ranking system enhanced the democratic bias of the input by $0.14$, making the output results more democratic ($TOB=0.52)$.
On the other hand, for queries related to republican candidates, on average, the ranking system increased the republican bias of the input by $0.18$---to the point that it changed the polarity of the output bias making the final results republican ($TOB=-0.11$), even though the average input bias had a democratic leaning.

This change of polarity of bias via the ranking system is more noticeable for some republican candidates. For instance, for Chris Christie, Jeb Bush and Lindsey Graham, while the input bias was positive, indicating that more democratic leaning users tweeted about them,
the ranking system switched the leaning of the output results to republican. These shifts in the bias caused by the ranking system (that can also result in a polarity change), exhibit the ranking systems power in altering the inherent bias of the input data.

\begin{table}
\center
\small
\begin{tabular}{| l || c | c | c |}
\hline
  & \multicolumn{3}{c|}{{\bf TRB of Ranking Strategies}} \\
\cline{2-4}
{\bf Query} & {\bf Twitter's}   &  {\bf Most }   &  {\bf Most} \\
				 &  {\bf Ranking}   &  {\bf Retweeted}   &  {\bf Favorited} \\
 				 &  {\bf }   &  {\bf First}   &  {\bf First} \\
\hline
\multicolumn{4}{|c|}{{\bf Queries Related to Democratic Candidates}} \\
\hline
Hillary Clinton & 0.18 & 0.33 & 0.25 \\
Bernie Sanders & 0.16 & 0.22 & 0.16 \\
Martin O'Malley & 0.07 & 0.001 & 0.1 \\
\hline
\multicolumn{4}{|c|}{{\bf Queries Related to Republican Candidates}} \\
\hline
Donald Trump & 0.10 & 0.06 & 0.09 \\
Ted Cruz & $-0.37$ & $-0.49$ & $-0.35$ \\
Marco Rubio & $-0.29$ & $-0.36$ & $-0.27$ \\
Ben Carson & 0.26 & 0.23 & 0.25 \\
Chris Christie & $-0.41$ & $-0.40$ & $-0.34$ \\
Jeb Bush & $-0.40$ & $-0.46$ & $-0.34$ \\
Rand Paul & $-0.19$ & $-0.25$ & $-0.17$ \\
Carly Fiorina & $-0.22$ & $-0.17$ & $-0.18$ \\
John Kasich & 0.04 & 0.04 & 0.11 \\
Mike Huckabee & 0.18 & 0.11 & 0.19 \\
Rick Santorum & $-0.22$ & $-0.34$ & $-0.16$ \\
Lindsey Graham & $-0.52$ & $-0.45$ & $-0.56$ \\
George Pataki & $-0.26$ & $-0.22$ & $-0.23$ \\
Jim Gilmore & $-0.24$ & $-0.22$ & $-0.21$ \\
\hline
\multicolumn{4}{|c|}{{\bf Queries related to democratic debate}} \\
\hline
democratic debate & 0.05 & 0.21 & 0.12 \\
dem debate & 0.23 & 0.22 & 0.22 \\
\#democraticdebate & 0.07 & 0.08 & 0.14 \\
\#demdebate & 0.01 & $-0.01$ & 0.01 \\
\hline
\multicolumn{4}{|c|}{{\bf Queries related to republican debate}} \\
\hline
republican debate & 0.26 & 0.274 & 0.268 \\
rep debate & $-0.09$ & $-0.09$ & $-0.09$ \\
\#republicandebate & 0.05 & 0.08 & 0.17 \\
\#gopdebate & $-0.06$ & $-0.06$ & $-0.02$ \\
\hline
\end{tabular}
\caption{{\bf Time averaged ranking bias for different ranking strategies: (i) Twitter's ranking (Twitter search ``top'' results), (ii) Most retweeted tweet first ranking, and (iii) Most favorited tweet first ranking. Here a bias value closer to $+1.0$ indicates democratic bias and a value closer to $-1.0$ indicates republican bias.}}
\label{table:diff_ranking_strategies}
\end{table}

The ranking of posts in social media search systems is influenced by a number of factors, including
the content's popularity (e.g., number of retweets or favorites), the author's popularity (e.g., number of followers
or lists), as well as the recency of the post.
To gain some insight into the extent to which different factors contribute to the overall observed bias, we construct rankings ``solely'' based on individual factors and examine the bias of these rankings.
Table~\ref{table:diff_ranking_strategies} shows the time averaged ranking bias of Twitter's ranking, as well as, the bias for two ranking strategies, each based on a single measure of posts' popularity,
namely the number of retweets and the number of favorites, respectively.

Table~\ref{table:diff_ranking_strategies} shows that the ranking biases of the three strategies are quite similar to each other, indicating that popularity of the post can explain much of the observed bias in Twitter's ranking.
For instance, for Chris Christie, Jeb Bush and Lindsey Graham, even though most of the users tweeting
about them are democratic leaning (as shown by positive $TIB$ values in Table~\ref{table:query-topic-bias}), most of the popular tweets
about them are made by republican leaning users (as shown by the negative $TRB$ values for popularity based ranking strategies in
Table~\ref{table:diff_ranking_strategies}).
However, in some cases like Martin O'Malley, John Kasich, democratic debate and \#republicandebate, the difference in the $TRB$ values between Twitter's ranking and the popularity based rankings indicate that, while popularity of posts is a factor that could explain a large part of the observed bias in Twitter search results, there are probably other factors that contribute to the overall bias of the search results.
Note that the above experiment is just a first step towards unpacking the influence of the different factors on the overall bias of search results in
a social media like Twitter, and we defer a more in-depth analysis for the future.

\subsection{RQ2b: The Collective Contribution of the Input Data and the Ranking System}
\vspace{0.2cm}

Given the impact of the input data and the influence of the ranking system on the bias of Twitter search results, we were curious to explore the dynamics between the input data and the ranking system when producing the output bias. While Table~\ref{table:query-topic-bias} shows a variance of output biases across the queries as a result of the interplay between the two, we particularly looked for the cases where the resulting output bias could affect a user's search experience noticeably. Below, we describe two of these cases by first analyzing the output bias and then exploring the input data's and the ranking system's contribution to the resulting output bias for each of these cases.

{\bf \textit{The Case of Popular Candidates}}

\textbf{Examining the Output Bias}: Comparing the output bias of the candidates in Table ~\ref{table:query-topic-bias}, we found a noticeable trend in the output bias of popular candidates compared to the rest of the candidates with the same political leaning -- {\it the search results for the more popular candidates have a higher bias towards the opposing perspective.}\footnote{The popularity of a candidate is estimated from the polling data obtained from~\cite{polling-data} for December 2015.}
 For instance, the top search results for Hillary Clinton, the most popular democratic candidate during this time, contained fewer democratic results ($TOB=0.21$) than the search results for other democratic candidates. Similarly, the search results for the query Donald Trump, the most popular republican candidate during this time, contained fewer republican results ($TOB=0.29$) when compared to the search results of other republican candidates.
 Therefore, if a user searches for popular candidate names in Twitter search, the top results have much higher bias toward the opposing perspective, while this is not as extreme for the less popular candidates.

In Figure~\ref{fig:avg-dem-bias-top-republic-candidates}, we plotted the output bias for the search results of republican candidates ranked by their popularity
to examine if there is a correlation between the popularity of a candidate and his output bias.\footnote{We did not perform this test for democratic candidates because there were only three of them.} The negative slope of the line of best fit in Figure~\ref{fig:avg-dem-bias-top-republic-candidates} supports the observation that the more popular a candidate is, the  more is the opposing perspective reflected in his or her top search results.

\begin{figure}[tb]
\center
\includegraphics[height=0.95\columnwidth,angle=-90]{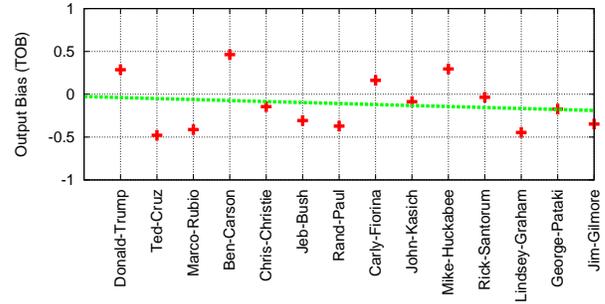}
\caption{{\bf The time averaged output bias $TOB$ in Twitter ``top'' search results for the Republican candidates -- candidates are listed left to right from highest to lowest popularity. }}
\label{fig:avg-dem-bias-top-republic-candidates}
\end{figure}

This situation may be undesirable for popular candidates, especially if users from the opposite perspective are more likely
to speak negatively about the candidate. We illustrate this case using Table~\ref{table:random_sample_hillary_trump}, which shows tweets randomly sampled from
the set of tweets posted by users with an opposing polarity when compared to the candidate. These tweets were included in the top search results for the queries Hillary Clinton and Donald Trump, and
they all either criticize or attempt to ridicule the candidates.
If an unbiased and potentially undecided voter searches for information about a popular presidential candidate, she would very likely see negative tweets posted by users of opposing leaning about the candidate. Similar to the findings of\mbox{\cite{search-engine-manipulation}}, this situation can be undesirable for the candidates as these negative tweets may impact the perceptions of these undecided voters.

\begin{table*}[tb]
\center
\small
\begin{tabular}{|  p{0.9\columnwidth} | p{0.9\columnwidth} |}
\hline
{\bf Randomly selected tweets from ``Hillary Clinton'' search results, which are posted by a republican leaning user}  & {\bf Randomly selected tweets from ``Donald Trump'' search results, which are posted by a democratic leaning user} \\
\hline
WT: Watchdog wants federal ethics probe of Clinton, possible improprieties http://bit.ly/1NvlrPA  &  Williamsburg, \#Brooklyn Dec 15 \#trump2016 \#MussoliniGrumpycat \#MakeAmericaHateAgain \#DonaldTrump @realDonaldTrump pic.twitter.com/Hj6DC7M7V1 \\
\hline
The Clintons both Bill and Hillary have a very long history of framing others while they commit the Crimes. History has destroyed the proof  &  Scotland defeats Trump on clean energy. Hopefully hell have a lot of time for golfing soon [url] \\
\hline
@CarlyFiorina: @realDonaldTrump is a big Christmas gift wrapped up under the tree for @HillaryClinton. [url]
&
Dirty little secret: Donald Trump is not a good debater.  \\
\hline
@CNN @HillaryClinton @BernieSanders hell no shes a murderer pic.twitter.com/zGQwR7dLZj  &  http://MLive.com - Where Donald Trumps Michigan campaign donations come from http://ow.ly/39hCWt \\
\hline
I dont care if youre a Democrat or Republican, how can you trust a word Hillary Clinton says and how can you consider voting for her??  &  Enjoy the sweet music of Donald Trump in Carol of the Trumps  [url] \\
\hline
\end{tabular}
\caption{{\bf Randomly selected tweets from the search results for the queries Hillary Clinton and Donald Trump, which are posted by a user with an opposite bias as compared to the candidate.}}
\label{table:random_sample_hillary_trump}
\end{table*}

{\bf The Contribution of Different Sources of Bias:}
Given the differing output bias trend for popular candidates compared to other candidates and the potential side effects, we explored if this differential output bias is inherent in the input tweets made by Twitter users or if it results from the ranking algorithm.
To understand this, we compared the input and output bias for the queries Hillary Clinton and Donald Trump. We found that the input bias for these queries leaned towards the opposite political view (similar to what we found in the output bias), indicating that the twitter users who talk about popular candidates are likely from the opposite political leaning than users who talk about the less popular candidates.

However, the manner in which the ranking system altered the input to produce the output and accordingly its bias was different for the two most popular candidates. While Hillary Clinton is discussed by republican users more than the other democratic candidates ($TIB=0.03$), the ranking system mitigated this undesirable situation by increasing the democratic output bias of the search results by a factor of seven as compared to the input bias. That is, the ranking system directed the search results for Hillary Clinton towards the perspective of her own party.
For Donald Trump the situation is the opposite. More democratic leaning users tweet about him ($TIB=0.19$) than the other republican candidates, and the ranking system enhanced this input bias resulting in more democratic tweets in the output search results. So, while the ranking system mitigated the opposite bias in the search results for Hillary Clinton, it enhanced it for Donald Trump.

This means that if a user searches for most popular candidates from each political party, the results favor Hillary Clinton over Donald Trump, while this was not the case in the input data.
These opposing dynamics between the input data and the ranking system, while inadvertent, can have serious implications, especially for the candidate for whom the ranking system enhanced the view points of the opposite leaning users.

{\bf \textit{Different Phrasings of Similar Queries}}

\textbf{Examining the Output Bias}: Different users who seek information about a certain topic might phrase their queries differently; e.g., for the event republican debate, the queries could be republican debate, rep debate, \#republicandebate or \#gopdebate.
While these differently phrased queries refer to the exact same event, the search results might differ substantially, particularly in the situation when users of different leanings selectively use different keywords and hashtags in their tweets to refer to the same event. This difference in search results for slightly differently phrased queries is common in many search engines, but whether the results for these similar queries exhibit different political biases is an open question.

To answer this question, in Table~\ref{table:query-topic-bias}, we compare the output biases of similarly phrased queries referring to the same event. We found that the output bias of two similar search queries for the same event can differ noticeably.
For example, the output results for the query republican debate have approximately twice the democratic leaning bias ($TOB=0.53$) as compared to the the output bias of the query rep debate ($TOB=0.31$). Similar noticeable differences also exist in the output bias for the queries related to the event democratic debate (e.g., democratic debate vs \#democraticdebate).

{\bf The Contribution of Different Sources of Bias:} To understand the sources of these differences between the output bias of similarly phrased queries, we compared the contributions of the input data and the ranking system to their output bias.
 As seen in Table~\ref{table:query-topic-bias}, the input data contributes more to the output bias for similarly phrased queries. In a few cases, however, the ranking system affects the input data noticeably, leading to two similarly phrased queries with similar input biases to have search results with very different output biases.
For example, comparison of the biases of the queries republican debate ($TIB=0.27$) and rep debate ($TIB=0.40$) reveals that while the input bias for these queries is similar, the ranking system altered their bias in opposing directions. That is, while the ranking system increased the bias of the republican debate query by $0.26$ making it more democratic, it decreased the bias of the rep debate query by $0.09$, making it more republican. Even when the input bias of these two queries was similar, the ranking system made one query significantly more democratic than the other. This example illustrates the influence a search system exerts on the input data, by curating search results with different biases for two similar queries with similar input biases.

These examples illustrate the interplay between the input data and the ranking system which produces different output biases for similarly phrased queries and for queries on popular political candidate. These observations lead to new questions for search engine design: How do these complex interactions between the input data and the ranking system affect the users' search experience and how can we make users aware of these effects? In the next section,
we briefly discuss solutions for signaling bias in search results to the users.

\section{Discussion}

\subsection{Generalizability of the Bias Quantification Framework}

\vspace{0.2cm}

{\bf Extending to other search engines:}
\noindent
Our proposed framework can be applied to other search engines (e.g., web search engines, other social media search engines) to quantify the bias in the search results even if the search algorithm hides within a black box and the internal details
of the retrieval and ranking algorithms are not known (which is almost always the case for real-world proprietary search engines).
The only pre-requisite for applying the framework is that a methodology for measuring the bias of
individual data items (e.g., web-pages, tweets) is available.

Our bias quantification framework (shown in
Figure~\ref{fig:seach_bias_framework}) works well for social media
search, where it is possible to separate the contributions of the
input data and the ranking algorithm to the final output
bias. However, in many modern-day IR systems, data items in the corpus
are directly ranked based on their relevance to the query,
without generating any intermediate set of relevant items. In such
systems,
it is hard to disentangle the bias introduced by the input data from
the bias added by the ranking process. But, we can still compare the
{\it relative biases} of different ranking systems, under the
assumption that both operate upon similar data corpora. For instance,
the ranking algorithms of Bing and Google search could be compared
with each other by observing their output biases for the same set of
queries.

{\bf Extending to multiple perspectives scenario:}
Even though we have applied our quantification framework for a two-perspective scenario of US politics,
our framework can be extended to multiple perspectives by having a bias vector for each item, rather than
a scalar score as we have now.
For instance, if the search scenario under consideration
has $p$ different perspectives associated with each query (such as queries related to a political election contested by $p$ political parties), then each data item (e.g., a tweet) can be represented as a $p$-dimensional bias vector.
Formally, the bias vector for the $i$-th data item would be given by $V_i$ = [$v_i^{1}$, $v_i^{2}$,...,
  $v_i^{p}$],
where $v_i^{j}$ gives a measure of how biased the $i$-th data item is along the
$j$-th perspective, with values in the range of $[-1, 1]$.
A value of $v_i^{j} = 1$ indicates that the item supports the $j$-th perspective,
$v_i^{j} = -1$ indicates that the item opposes the perspective, whereas
$v_i^{j} = 0$ indicates that the item is neutral with respect to that perspective.
Then Equations 2 to 5 can be converted to their vector addition formulations, to
measure the input, output and ranking biases.
A challenging aspect of extending to a $p$-dimensional scenario
would be to develop a methodology to capture these bias vectors, and
it would be interesting to explore this in the future.

\subsection{Signaling Political Bias in Search Results}
\vspace{0.2cm}

While our analyses show that social media search results have varying amounts of political bias, how this bias can be
tackled is still an open question.
In this section, we briefly discuss some potential solutions to this question, but their in-depth exploration is left for future work.

{\bf Incorporating Bias into the Ranking System:}
One solution for controlling bias could be to develop a ranking mechanism that considers bias as a metric and
trades-off relevance and bias of the search results.
For instance, a minimal value of average bias of the search
results could be achieved by interleaving results
from the various perspectives of a search query, using
methods similar to those proposed to inject diversity
in search results~\cite{search-engine-diversity,Welch-diversity-info-queries}.
However, finding a suitable equilibrium between
bias and other ranking factors is challenging,
specially because the trade-off point is likely to be domain and user specific.
Additionally, including bias as a factor in the ranking systems
might lead to a degradation of the top search results along
relevancy, popularity, recency, or other metrics.

{\bf Making bias transparent:}
Given the aforementioned problems with changing the ranking of the search results, an
alternative method of addressing the bias can be to incorporate the bias into
the front-end of the search engine (by visualizing the bias of search result), rather than into the ranking algorithm itself.
Through this method, while the search algorithm's efficiency is not compromised,
users can be made aware of the possible biases in their search results by marking each search result with its political bias.
Such a nudging practice has been used widely in the literature
for purposes like delivering multiple aspects of news in social media~\cite{park2009newscube} and
encouraging reading of diverse political opinions~\cite{munson2010presenting,munson2013encouraging}.

A hybrid approach of the above two methods could also be proposed,
which not only shows the bias of each search result, but also separates the results of the two political
perspectives (republican and democratic) and shows them as {\it distinct ranked lists},
with each distinct list retaining the ranking of the results in the original ranked list.
By preserving the original search engine's ranking within each list,
this methodology ensures that the quality of the top search
results does not degrade across other metrics
such as relevance, popularity, and recency.

Developing tools to signal political bias in search results, and conducting
user studies to understand how users
interact with such alternative search interface designs are important, and are left for future work.

\subsection{Auditing Black Boxes}
\vspace{0.2cm}
Recently, the rise of algorithmic platforms' influence on users' online experience has motivated many studies to audit these platforms and understand their biases. While some of these algorithmic systems' functionalities are open to the public, making the auditing process easier, most of them are not. The walls of intellectual proprietary, high complexity of these algorithms
and the perils of gaming a system via malicious users put these algorithms in a black box, making it almost infeasible to
have access to an algorithm's specifications from outside, like in our study. While we know about a few general factors that a search engines takes into account in curating the search results (such as relevancy, popularity, and recency), there are hundreds of other features that are hidden in a blackbox, preventing us as researchers from being able to pin-point the exact feature(s) of the algorithm which might be leading to the bias being introduced in the search results.

Therefore, building on previous studies that have adopted this ``black box'' view for an algorithmic system while auditing it \cite{eslami2015always,liao2016snowden,hannak2014measuring,web-search-personalization,chen2015peeking}, we characterized the bias of the ranking algorithm in Twitter's search platform, without knowing its internals. Our proposed auditing framework can help users, system designers and
researchers to become aware of possible biases of a search process, while they might not be aware of the details of the process itself.
For users, this awareness can result in more intelligent use of a system, knowing that their search results
can be far from neutral in some cases. For system designers, such auditing platforms can be used to
investigate the algorithm's specifications, particularly when the bias has been introduced by the algorithm and not the system input.
And finally, researchers and watchdog organizations can actively utilize such auditing platforms
to measure bias and compare it among different search platforms, making the research community and the system designers aware of potentially misleading biases.

\subsection{Distinguishing the Sources of Bias: From Development to Design}
\vspace{0.2cm}

Our study has proposed a first step towards distinguishing the different sources of bias in a search engine. The observation that a significant part of the bias comes from the input data in some cases (without the interference of the ranking system) highlights the need to add data-driven perspective to algorithm audit studies, such that not only the output data is observed to understand the algorithm's discriminatory or biased behavior, but also the potential biases in the input data are investigated.

Although a part of the output bias stems from the input data, our study revealed that an algorithm can influence the inherent bias of the input data significantly, to the extent of even changing its polarity.
This algorithmic power is of great importance particularly when a design choice in the interface gives higher visibility to the algorithm's results than the input data. Twitter search, for example, provides users with  ``top'' and ``live'' search results, where the former is the ranked list of tweets output by the search system
and the latter is simply all the tweets containing the search query in reverse chronological order, i.e. the input data. This means that a user looking for information about a topic is able to see both the input data (which might be more representative of the full Twitter data) and the output data in Twitter search if she wants. However, to give people more relevant results in a shorter amount of time, Twitter has made the top results (the ranking systems's output) the default option rather than the live results (the input data). This design choice, while legitimate, gives the ranked results more visibility, making the output bias and accordingly the ranking system's influence more highlighted than the input bias in the design.
These design choices, along with the significant effect of ranking system on the final output search bias, suggest a new avenue of research for studying when and how the influence of the ranking system (particularly when it is more visible in the design), might result in showing a totally different side of a story to a user than what the input data would show to her.

\section{Limitations}

In this study, we focussed on a limited set of queries that were either related to a political event or a political candidate, due to the limitation on the number of queries we could submit via Twitter API. Given the variation of bias over different queries, we believe extending our query set to more general political queries on controversial topics such as gun control and abortion would strengthen our analysis, and we hope to pursue this in the  future.
Another limiting factor in our study was using the simplifying assumption of considering a user as either neutral, pro-democratic or pro-republican. Under this assumption we can not have a user who is partially both pro-republican and pro-democrat. However, we should clarify that for doing this classification, we still considered two scores for each user, one the similarity to republicans and the other similarity to democrats. Currently to give the user a final leaning, we consider the difference of these similarities. However, in the future, we can use these two similarities to determine the extent to which a user is pro-democratic as well as pro-republican to have a more nuanced view of political leanings of users.
And lastly, while we have discussed some potential solutions for signaling political bias in search results, we have not as yet implemented our proposed solutions and evaluated the effect of this signaling on the users' search experience. This exploration would be a good follow up of our current study.

\section{Conclusion}  \label{sec:conclu}

\noindent
To our knowledge, the present study developed the first framework to quantify bias of ranked results in a search process, while being able to distinguish between different sources of bias. Via this ability, our framework not only measures the bias in the output ranked list of search results, but is also able to capture how much of this bias is due to the biased set of input data to the ranking system and how much is contributed by the ranking system itself. In the earlier search engine bias studies, these two factors have not been separated out. Our analyses revealed the significant effect of both input data and the ranking algorithm in producing considerable bias in Twitter search results based on various factors such as the topic of a query or how the query is phrased.

As more and more users are relying on social media search
to follow live events and news on personalities~\cite{Teevan_TwitterSearch},
the varying biases in search results can have significant
impact on the impression that users form about the different events
and personalities~\cite{google-we-trust,search-engine-manipulation}.
We end by calling for mechanisms to make users more aware of the potential
biases in search results, e.g., by presenting the results in a way that makes
the different perspectives transparent to the user.
Giving the users more control over the bias beyond just making them more aware of its existence could be a potential next step.
For example, users could be given controls or filters to re-rank the search results to adjust for search biases -- just as
they can in ranking of products and services on sites like Amazon or  Yelp.

\balance{}

\bibliographystyle{SIGCHI-Reference-Format}

\begin{thebibliography}{00}


\ifx \showCODEN    \undefined \def \showCODEN     #1{\unskip}     \fi
\ifx \showDOI      \undefined \def \showDOI       #1{{\tt DOI:}\penalty0{#1}\ }
  \fi
\ifx \showISBNx    \undefined \def \showISBNx     #1{\unskip}     \fi
\ifx \showISBNxiii \undefined \def \showISBNxiii  #1{\unskip}     \fi
\ifx \showISSN     \undefined \def \showISSN      #1{\unskip}     \fi
\ifx \showLCCN     \undefined \def \showLCCN      #1{\unskip}     \fi
\ifx \shownote     \undefined \def \shownote      #1{#1}          \fi
\ifx \showarticletitle \undefined \def \showarticletitle #1{#1}   \fi
\ifx \showURL      \undefined \def \showURL       #1{#1}          \fi

\bibitem{Adamic-blogosphere}
{Lada~A. Adamic} {and} {Natalie Glance}. 2005.
\newblock \showarticletitle{The Political Blogosphere and the 2004 U.S.
  Election: Divided They Blog}. In {\em Proc. LinkKDD}.
\newblock


\bibitem{barocas2014big}
{Solon Barocas} {and} {Andrew~D Selbst}. 2014.
\newblock \showarticletitle{Big data's disparate impact}.
\newblock {\em Available at SSRN 2477899\/} (2014).
\newblock


\bibitem{user-interests-recsys}
{Parantapa Bhattacharya}, {Muhammad~Bilal Zafar}, {Niloy Ganguly}, {Saptarshi
  Ghosh}, {and} {Krishna~P. Gummadi}. 2014.
\newblock \showarticletitle{{Inferring User Interests in the Twitter Social
  Network}}. In {\em {Proc. ACM RecSys}}.
\newblock


\bibitem{chen2015peeking}
{Le Chen}, {Alan Mislove}, {and} {Christo Wilson}. 2015.
\newblock \showarticletitle{Peeking beneath the hood of uber}. In {\em In Proc.
  of the 2015 ACM Conference on Internet Measurement Conference}. ACM,
  495--508.
\newblock


\bibitem{political-alignment-twitter-socialcom}
{M. Conover}, {B. Gon\c{c}alves}, {J. Ratkiewicz}, {A. Flammini}, {and} {F.
  Menczer}. 2011a.
\newblock \showarticletitle{Predicting the Political Alignment of Twitter
  Users}. In {\em Proc. IEEE SocialCom}.
\newblock


\bibitem{political-polarization-twitter-icwsm}
{M. Conover}, {J. Ratkiewicz}, {Matthew Francisco}, {B Gon\c{c}alves}, {F.
  Menczer}, {and} {A. Flammini}. 2011b.
\newblock \showarticletitle{Political Polarization on Twitter}. In {\em Proc.
  AAAI ICWSM}.
\newblock


\bibitem{datta2014automated}
{Amit Datta}, {Michael~Carl Tschantz}, {and} {Anupam Datta}. 2014.
\newblock \showarticletitle{Automated Experiments on Ad Privacy Settings: A
  Tale of Opacity, Choice, and Discrimination}.
\newblock {\em Choice, and Discrimination. arXiv. org\/} (2014).
\newblock


\bibitem{search-engine-manipulation}
{Robert Epstein} {and} {Ronald~E. Robertson}. 2015.
\newblock \showarticletitle{{The search engine manipulation effect (SEME) and
  its possible impact on the outcomes of elections}}.
\newblock {\em {Proc. of the National Academy of Sciences (PNAS)}\/} {112}, 33
  (2015), E4512--E4521.
\newblock


\bibitem{eslami2015always}
{Motahhare Eslami}, {Aimee Rickman}, {Kristen Vaccaro}, {Amirhossein Aleyasen},
  {Andy Vuong}, {Karrie Karahalios}, {Kevin Hamilton}, {and} {Christian
  Sandvig}. 2015.
\newblock \showarticletitle{I always assumed that I wasn't really that close to
  [her]: Reasoning about Invisible Algorithms in News Feeds}. In {\em In Proc.
  of the 33rd Annual ACM Conference on Human Factors in Computing Systems}.
  ACM, 153--162.
\newblock


\bibitem{wh-study-on-biases}
{USA Executive Office of~the President}. 2016.
\newblock {Big Data: A Report on Algorithmic Systems, Opportunity,and Civil
  Rights}.
\newblock {\it http://tinyurl.com/Big-Data-White-House}.   (2016).
\newblock


\bibitem{search-engine-bias-mitigation}
{S. Fortunato}, {A. Flammini}, {F. Menczer}, {and} {A. Vespignani}. 2006.
\newblock \showarticletitle{{Topical interests and the mitigation of search
  engine bias}}.
\newblock {\em {Proc. of the National Academy of Sciences (PNAS)}\/} {103}, 34
  (2006), 12684--12689.
\newblock


\bibitem{cognos_sigir}
{Saptarshi Ghosh}, {Naveen Sharma}, {Fabricio Benevenuto}, {Niloy Ganguly},
  {and} {Krishna Gummadi}. 2012.
\newblock \showarticletitle{{Cognos: Crowdsourcing Search for Topic Experts in
  Microblogs}}. In {\em {Proc. ACM SIGIR}}.
\newblock
\showISBNx{978-1-4503-1472-5}


\bibitem{political-preference-twitter-sigchi}
{Jennifer Golbeck} {and} {Derek Hansen}. 2011.
\newblock \showarticletitle{{Computing Political Preference Among Twitter
  Followers}}. In {\em {ACM SIGCHI}}.
\newblock


\bibitem{web-search-personalization}
{Aniko Hannak}, {Piotr Sapiezynski}, {Arash Molavi~Kakhki}, {Balachander
  Krishnamurthy}, {David Lazer}, {Alan Mislove}, {and} {Christo Wilson}. 2013.
\newblock \showarticletitle{Measuring Personalization of Web Search}. In {\em
  Proc. WWW}.
\newblock


\bibitem{hannak2014measuring}
{Aniko Hannak}, {Gary Soeller}, {David Lazer}, {Alan Mislove}, {and} {Christo
  Wilson}. 2014.
\newblock \showarticletitle{Measuring price discrimination and steering on
  e-commerce web sites}. In {\em In Proc. of the 2014 conference on internet
  measurement conference}. ACM, 305--318.
\newblock


\bibitem{flicker-autotag}
{Alex Hern}. 2015.
\newblock {Flickr faces complaints over `offensive' auto-tagging for photos}.
\newblock {\it http://tinyurl.com/Flickr-AutoTagging}.   (2015).
\newblock


\bibitem{himelboim2013birds}
{Itai Himelboim}, {Stephen McCreery}, {and} {Marc Smith}. 2013.
\newblock \showarticletitle{Birds of a feather tweet together: Integrating
  network and content analyses to examine cross-ideology exposure on Twitter}.
\newblock {\em Journal of Computer-Mediated Communication\/} {18}, 2 (2013),
  40--60.
\newblock


\bibitem{web-search-personalization-geolocation}
{Chloe Kliman-Silver}, {Aniko Hannak}, {David Lazer}, {Christo Wilson}, {and}
  {Alan Mislove}. 2015.
\newblock \showarticletitle{Location, Location, Location: The Impact of
  Geolocation on Web Search Personalization}. In {\em Proc. ACM IMC}.
\newblock


\bibitem{Koutra-events-controversies}
{Danai Koutra}, {Paul~N. Bennett}, {and} {Eric Horvitz}. 2015.
\newblock \showarticletitle{Events and Controversies: Influences of a Shocking
  News Event on Information Seeking}. In {\em {Proc. WWW}}.
\newblock


\bibitem{liao2016snowden}
{Q~Vera Liao}, {Wai-Tat Fu}, {and} {Markus Strohmaier}. 2016.
\newblock \showarticletitle{\# Snowden: Understanding Biases Introduced by
  Behavioral Differences of Opinion Groups on Social Media}. In {\em In Proc.
  of the 2016 CHI Conference on Human Factors in Computing Systems}. ACM,
  3352--3363.
\newblock


\bibitem{neiman_news}
{Joseph Lichterman}. 2010.
\newblock {New Pew data: More Americans are getting news on Facebook and
  Twitter}.
\newblock   (2010).
\newblock
\newblock
\shownote{{\it http://tinyurl.com/News-on-Social-Media}.}


\bibitem{liu2014twitter}
{Zhe Liu} {and} {Ingmar Weber}. 2014.
\newblock \showarticletitle{Is Twitter a public sphere for online conflicts? A
  cross-ideological and cross-hierarchical look}. In {\em International
  Conference on Social Informatics}. Springer, 336--347.
\newblock


\bibitem{political-preference-twitter-asonam}
{Aibek Makazhanov} {and} {Davood Rafiei}. 2013.
\newblock \showarticletitle{Predicting Political Preference of Twitter Users}.
  In {\em Proc. ASONAM}.
\newblock


\bibitem{irbook-manning}
{Christopher~D. Manning}, {Prabhakar Raghavan}, {and} {Hinrich Schutze}. 2008.
\newblock {\em Introduction to Information Retrieval}.
\newblock Cambridge University Press.
\newblock


\bibitem{twitter-pew}
{Amy Mitchell} {and} {Dana Page}. 2015.
\newblock \showarticletitle{The Evolving Role of News on Twitter and Facebook}.
\newblock {\em Pew Research Center\/} (2015).
\newblock


\bibitem{measure-search-engine-bias}
{Abbe Mowshowitz} {and} {Akira Kawaguchi}. 2005.
\newblock \showarticletitle{{Measuring search engine bias}}.
\newblock {\em {Information Processing and Management}\/} {41}, 5 (2005),
  1193--1205.
\newblock


\bibitem{munson2013encouraging}
{Sean~A Munson}, {Stephanie~Y Lee}, {and} {Paul Resnick}. 2013.
\newblock \showarticletitle{Encouraging Reading of Diverse Political Viewpoints
  with a Browser Widget.}. In {\em ICWSM}.
\newblock


\bibitem{munson2010presenting}
{Sean~A Munson} {and} {Paul Resnick}. 2010.
\newblock \showarticletitle{Presenting diverse political opinions: how and how
  much}. In {\em In Proc. of the SIGCHI conference on human factors in
  computing systems}. ACM, 1457--1466.
\newblock


\bibitem{google-we-trust}
{Bing Pan}, {Helene Hembrooke}, {Thorsten Joachims}, {Lori Lorigo}, {Geri Gay},
  {and} {Laura Granka}. 2007.
\newblock \showarticletitle{In Google We Trust: Users' Decisions on Rank,
  Position, and Relevance}.
\newblock {\em {Journal of Computer-Mediated Communication}\/}  {12} (2007),
  801--823.
\newblock
Issue 3.


\bibitem{park2009newscube}
{Souneil Park}, {Seungwoo Kang}, {Sangyoung Chung}, {and} {Junehwa Song}. 2009.
\newblock \showarticletitle{{NewsCube: delivering multiple aspects of news to
  mitigate media bias}}. In {\em {Proc. ACM CHI}}.
\newblock


\bibitem{pew-twitter-dem}
Pew Research 2013.
\newblock {Twitter Reaction to Events Often at Odds with Overall Public
  Opinion}.
\newblock {\it
  http://www.pewresearch.org/2013/03/04/twitter-reaction-to-events-often-at-odds-with-overall-public-opinion/}.
    (2013).
\newblock


\bibitem{twitter-language-leaning-plosone}
{Matthew Purver} {and} {Sylwester Karolina}. 2015.
\newblock \showarticletitle{{Twitter Language Use Reflects Psychological
  Differences between Democrats and Republicans}}.
\newblock {\em {PLoS ONE}\/} {10}, 9 (2015), e0137422.
\newblock


\bibitem{polling-data}
Real Clear Politics 2015.
\newblock {RealClearPolitics -- Election 2016 -- 2016 Republican Presidential
  Nomination}.
\newblock {\it http://tinyurl.com/us-republican-polling-data}.   (2015).
\newblock


\bibitem{sandvig2014auditing}
{Christian Sandvig}, {Kevin Hamilton}, {Karrie Karahalios}, {and} {Cedric
  Langbort}. 2014.
\newblock \showarticletitle{Auditing algorithms: Research methods for detecting
  discrimination on internet platforms}.
\newblock {\em Data and Discrimination: Converting Critical Concerns into
  Productive Inquiry\/} (2014).
\newblock


\bibitem{semaan2014social}
{Bryan~C Semaan}, {Scott~P Robertson}, {Sara Douglas}, {and} {Misa Maruyama}.
  2014.
\newblock \showarticletitle{Social media supporting political deliberation
  across multiple public spheres: towards depolarization}. In {\em In Proc. of
  the 17th ACM conference on Computer supported cooperative work \& social
  computing}. ACM, 1409--1421.
\newblock


\bibitem{whoiswho_wosn}
{Naveen Sharma}, {Saptarshi Ghosh}, {Fabricio Benevenuto}, {Niloy Ganguly},
  {and} {Krishna Gummadi}. 2012.
\newblock \showarticletitle{{Inferring Who-is-Who in the Twitter Social
  Network}}. In {\em {Proc. ACM WOSN}}.
\newblock


\bibitem{smith2013role}
{Laura~M Smith}, {Linhong Zhu}, {Kristina Lerman}, {and} {Zornitsa Kozareva}.
  2013.
\newblock \showarticletitle{The role of social media in the discussion of
  controversial topics}. In {\em Social Computing (SocialCom), 2013
  International Conference on}. IEEE, 236--243.
\newblock


\bibitem{sweeney2013discrimination}
{Latanya Sweeney}. 2013.
\newblock \showarticletitle{Discrimination in online ad delivery}.
\newblock {\em Queue\/} {11}, 3 (2013), 10.
\newblock


\bibitem{sep-ethics-search}
{Herman Tavani}. 2014.
\newblock \showarticletitle{Search Engines and Ethics}.
\newblock In {\em The Stanford Encyclopedia of Philosophy} (spring 2014 ed.),
  {Edward~N. Zalta} (Ed.).
\newblock


\bibitem{Teevan_TwitterSearch}
{Jaime Teevan}, {Daniel Ramage}, {and} {Merredith~Ringel Morris}. 2011.
\newblock \showarticletitle{\#TwitterSearch: a comparison of microblog search
  and web search}. In {\em {Proc. ACM WSDM}}.
\newblock


\bibitem{search-engine-bias-thesis}
{Elizabeth Van~Couvering}. 2010.
\newblock {\em {Search engine bias: the structuration of traffic on the
  World-Wide Web}}.
\newblock Ph.D. Dissertation. The London School of Economics and Political
  Science.
\newblock


\bibitem{Vaughan-search-coverage-bias}
{Liwen Vaughan} {and} {Mike Thelwall}. 2004.
\newblock \showarticletitle{Search Engine Coverage Bias: Evidence and Possible
  Causes}.
\newblock {\em Information Processing and Management\/} {40}, 4 (May 2004),
  693--707.
\newblock


\bibitem{Weber-query-logs}
{Ingmar Weber}, {Venkata Rama~Kiran Garimella}, {and} {Erik Borra}. 2012.
\newblock \showarticletitle{{Mining Web Query Logs to Analyze Political
  Issues}}. In {\em Proc. ACM WebSci}.
\newblock


\bibitem{weber2013political}
{Ingmar Weber}, {Venkata Rama~Kiran Garimella}, {and} {Asmelash Teka}. 2013.
\newblock \showarticletitle{Political hashtag trends}. In {\em European
  Conference on Information Retrieval}. Springer, 857--860.
\newblock


\bibitem{Welch-diversity-info-queries}
{Michael~J. Welch}, {Junghoo Cho}, {and} {Christopher Olston}. 2011.
\newblock \showarticletitle{Search Result Diversity for Informational Queries}.
  In {\em Proc. WWW}.
\newblock


\bibitem{biased-language-naacl}
{Tae Yano}, {Philip Resnik}, {and} {Noah~A. Smith}. 2010.
\newblock \showarticletitle{Shedding (a Thousand Points of) Light on Biased
  Language}. In {\em Proc NAACL HLT Workshop on Creating Speech and Language
  Data with Amazon's Mechanical Turk (CSLDAMT)}.
\newblock


\bibitem{yardi2010dynamic}
{Sarita Yardi} {and} {Danah Boyd}. 2010.
\newblock \showarticletitle{Dynamic debates: An analysis of group polarization
  over time on twitter}.
\newblock {\em Bulletin of Science, Technology \& Society\/} {30}, 5 (2010),
  316--327.
\newblock


\bibitem{average-prec-incomplete-judgement-cikm}
{Emine Yilmaz} {and} {Javed~A. Aslam}. 2006.
\newblock \showarticletitle{Estimating Average Precision with Incomplete and
  Imperfect Judgments}. In {\em Proc. ACM CIKM}.
\newblock


\bibitem{search-engine-diversity}
{Elad Yom-Tov}, {Susan Dumais}, {and} {Qi Guo}. 2013.
\newblock \showarticletitle{{Promoting Civil Discourse Through Search Engine
  Diversity}}.
\newblock {\em {Social Science Computer Review}\/}  {32} (2013), 145--154.
\newblock
Issue 2.


\bibitem{impartiality-icwsm}
{Muhammad~Bilal Zafar}, {Krishna~P. Gummadi}, {and} {Cristian
  Danescu-Niculescu-Mizil}. 2016.
\newblock \showarticletitle{Message Impartiality in Social Media Discussions}.
  In {\em Proc. AAAI ICWSM}.
\newblock


\bibitem{news-political-leaning-icwsm}
{Daniel~Xiaodan Zhou}, {Paul Resnick}, {and} {Qiaozhu Mei}. 2011.
\newblock \showarticletitle{Classifying the Political Leaning of News Articles
  and Users from User Votes}. In {\em Proc. AAAI ICWSM}.
\newblock


\end{thebibliography}

\end{document}